\documentclass[journal]{IEEEtran}
\usepackage{graphicx}
\usepackage{color}
\usepackage{amsmath}
\usepackage{subfigure}
\usepackage{amssymb}
\usepackage{amsfonts}
\usepackage{psfrag}
\usepackage{amsmath,epsfig,cite,amsfonts,amssymb,psfrag,subfig}
\usepackage{cite}
\usepackage{algorithm}
\usepackage{algorithmic}
\usepackage{balance}

\ifCLASSINFOpdf
\else
\fi

\begin{document}
%
\title{Rateless Codes for the Multi-Way Relay Channel}

\author{\IEEEauthorblockN{Iqbal Hussain, Ming Xiao, and Lars K. Rasmussen}\\
\IEEEauthorblockA{School of Electrical Engineering, Communication Theory Laboratory \\
KTH Royal Institute of Technology and the ACCESS Linnaeus Center \\
Stockholm, Sweden, Email: \{iqbalh,mingx,lkra\}@kth.se}
}
\maketitle

\begin{abstract}
We consider distributed Luby transform (DLT) codes for efficient packet transmission in a multi-way relay network, where the links are modeled as erasure channels. Density evolution is applied for asymptotic performance analysis, and subsequently used in a linear-programming design framework for optimizing the degree distribution at the relay in terms of overhead. Moreover a buffer is introduced at the relay to enable efficient downlink transmission even if packets are lost during uplink transmission. Performance losses in terms of delay and/or erasure rates caused by link erasures during uplink transmission are thus alleviated. The proposed DLT codes provide significant improvements in overhead and decoded erasure rates. Numerical results for finite-length codes follow closely the asymptotic analysis. Our results demonstrate that the proposed buffer-based DLT codes outperform its counterparts for lossy uplink transmission. 

\end{abstract}

%
\IEEEpeerreviewmaketitle
\section{Introduction}
\IEEEPARstart
{S}{ince} the seminal work of \cite{CG79}, relay techniques have received significant attention in the research community. The two-way relay channel has been studied in detail in \cite{WNPS10} and the references therein. As a generalization 
the multi-way relay (MR) channel has been investigated in \cite{GunYen09isit}. Practical examples of the MR channel include information exchange between multiple sensor/actuator nodes via an access point in a networked control system, as well as tele-conferencing in cellular networks. In the case where channel state information (CSI) is known to all the users and the relay, the MR channel was studied for Gaussian channels in \cite{OKJ10} and for binary symmetric channels in \cite{JOK10}.

When no CSI is available at the transmitter, a promising technique for time-varying channels is to employ rateless codes. The family of Luby transform (LT) codes \cite{LT} represents the first practical realization of such codes. The original LT codes \cite{LT} exhibit excellent performance over a binary erasure channel (BEC) but encoding/decoding complexities are high especially at large block lengths. Raptor codes \cite{Sh06} were proposed for rateless codes to achieve linear encoding/decoding complexities. Raptor codes have been shown to mitigate the error floor over noisy channels \cite{ES06}. When applied to relay channels, a novel rateless coding framework is proposed for a three node network in \cite{CY07}.  Distributed LT (DLT) codes were introduced for a relay network with multiple sources in \cite{PKF07}, where the relay node selectively combines incoming symbols from the sources and forwards the output to the destination. The degree distributions at the sources and the combining operation at the relay were coordinated to obtain a Soliton-like degree distribution at the destination. A more general approach to DLT codes has been proposed in \cite{SPD09} where the relay is allowed to perform re-encoding rather than simple XORing of incoming symbols from the sources. These 
codes 
outperform the  DLT codes in \cite{PKF07}; however, for lossy uplink transmission, they 
exhibit a higher delay, which is not desirable in 
latency-limited applications. 
The design approach of these conventional DLT codes cannot be extended to MR channels due to the presence of multiple destinations.

Here we extend the use of DLT codes to the erasure MR (EMR) channel. Wireless erasure networks, as thoroughly investigated in \cite{DGPHE06} are of particular interest for two reasons: Firstly they are a meaningful abstractions of real-world systems; and secondly they form a class of networks which, in general, allows for mathematically tractable information-theoretic problem formulations. The erasure channel models the fundamental detrimental effects of deep fade events experienced over a wireless channel.

In particular we formulate a design framework for optimizing the degree distribution at the relay, with the aim of increasing the transmission efficiency. We use density evolution to determine the asymptotic performance, which is subsequently exploited in our proposed design framework. A crucial component in our scheme is the introduction of a buffer at the relay in contrast to the conventional DLT coding schemes. The purpose of the buffer is to ensure that a full set of input symbols is always available at the relay. It follows that the relay can transmit according to the designed degree distribution in the downlink even if packets are lost during the current uplink transmission.
\section{Preliminaries}
The family of LT codes is a class of sparse graph codes designed for erasure channels. The rateless encoder maps a sequence of $K$ information symbols to a potentially unlimited stream of coded symbols where the symbol cardinality can be arbitrary. For simplicity, we consider binary symbols but the results can easily be extended to arbitrary packet sizes by bitwise modulo-2 operations. An information bit is also denoted as a variable node, and a coded bit is termed a check node in the code graph. Decoding is performed using an iterative message-passing decoder operating on the graph generated by the correctly received coded bits \cite{LT}. Once the decoder has recovered all the information bits an acknowledgment is forwarded to the transmitter to halt transmission.

The degree of a check node is the number of variable nodes involved in determining the value of the check node. The degree for a particular instance is determined by sampling the check-node degree distribution, represented by $\Omega(x)=\sum_{j=1}^{J}\Omega_j x^j$. Here $\Omega_j$ denotes the probability of choosing a degree $j$ check node, while $J$ denotes the maximum check-node degree. As the variable nodes are selected uniformly at random by the check nodes, the variable-node degrees are binomial distributed. The binomial distribution is well approximated by a Poisson distribution \cite{Sh06}, and consequently, the variable-node degree distribution can be approximated by
\begin{equation}\label{eqn:vardeg}
\Lambda(x) = \sum_{i=1}^{I} \Lambda_i x^i \approx e^{\mu(x-1)}.
\end{equation}
Here $I$ is the maximum variable-node degree, and $\Lambda_{i}$ is the probability of a variable node of degree $i$ computed as
\begin{equation}\label{eqn:vardeg1}
\Lambda_{i}\approx\frac{e^{-\mu }\mu^{i}}{i!}.
\end{equation}
The Poisson parameter $\mu$ is the average degree of a variable node at the output of the encoder. For the asymptotic performance analysis, the edge-perspective degree distributions, denoted by $\omega(x)$ and $\lambda(x)$ for check nodes and variable nodes, respectively, are more convenient. The relationships between node-perspective and edge-perspective degree distributions are evaluated as follows \cite{Sh06}
\begin{align}\label{eqn:vnedgedeg}
\omega(x)&=\Omega'(x)/\Omega'(1)=\sum_{j=1}^{J}\omega_jx^{j-1},\\
\lambda(x)&=\Lambda'(x)/\Lambda'(1)=\sum_{i=1}^{I}\lambda_ix^{i-1}\approx e^{\mu(x-1)},
\end{align}
where $f'(x)$ is the derivative of $f(x)$ with respect to $x$.
\begin{figure}[!t]
\centering
    \psfrag{R}[][][0.7]{Relay}
    \psfrag{u1}[][][0.7]{$1$}
    \psfrag{u2}[][][0.7]{$2$}
    \psfrag{ldot}[][][1]{$\rotatebox{90}{\ldots}$}
    \psfrag{ur}[][][0.7]{$r$}
    \psfrag{C1}[][][0.7]{$\color{blue}c_1$}
    \psfrag{C2}[][][0.7]{$\color{blue}c_2$}
    \psfrag{Cr}[][][0.7]{$\color{blue}c_r$}
    \psfrag{xj}[][][0.7]{$\color{red}x_j$}
    \includegraphics[width=33mm]{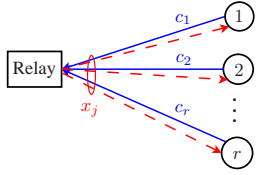}
    \caption{Erasure Multi-way Relay Channel; uplink phase denoted by solid lines; broadcast phase denoted by dotted lines.}
    \label{system_model}
\end{figure}
Consider the system model illustrated in Fig.~\ref{system_model}, where $r$ users intend to exchange information over the EMR channel. The uplink channel from user $i$ to the relay is a BEC with erasure probability $\epsilon_{u_i}$. Similarly, the downlink channel from the relay to user $i$ is a BEC with erasure probability $\epsilon_{d_i}$. We assume that no channel state information is available at any transmitter in the network. Each user maps $K$ information bits into a sequence of coded bits using a common degree distribution $\Omega(x)$. As shown in Fig.~\ref{system_model}, the transmission of each coded bit from users to destinations is divided into two phases. In the uplink phase time division is applied such that each user transmits one coded bit to the relay in one time slot. Thus, the uplink phase has $r$ such time slots. The coded bit transmitted from user $i$ is denoted by $c_{i}$ at a given uplink phase, and becomes a relay variable node at the relay. In the broadcast phase, the relay forwards a relay-coded bit to all the users. This bit is generated using a predetermined relay check-node degree distribution, denoted by $\Gamma(x)=\sum_{d=1}^{D} \Gamma_d x^d$. Here $\Gamma_d$ is the probability of choosing a degree $d$ relay check node, and $D$ is the maximum degree subject to the constraint $D\leq r$. At the end of the uplink phase, the relay samples a degree $d$ from the relay check-node degree distribution $\Gamma(x)$, followed by selecting, uniformly at random, $d$ users. 
The relay-coded bit is then generated by modulo-2 addition of the $d$ relay variable nodes. The relay-coded bit in the $j$-th ($j = 1, 2, \cdots$) broadcast phase is determined as
\vspace{-3mm}
\begin{equation}\label{RCB}
x_j=\overset{D}{\underset{i=1}{\bigoplus}}\psi_i \alpha_i c_i,
\end{equation}
where $\bigoplus$ denotes a modulo-2 sum. Here, $\psi_i=1$ if user $i$ is selected and $\psi_i=0$ otherwise. Furthermore, $\alpha_i=1$ with probability $(1-\epsilon_{u_i})$ and $\alpha_i=0$ with probability $\epsilon_{u_i}$ 
whether $c_i$ is erased in a given uplink phase or not. The relay broadcasts the relay-coded bit $x_j$ to the users in the broadcast phase. After receiving a relay-coded bit, each user 
removes its information bit from $x_j$ by simple modulo-2 addition. On receiving a sufficient number of relay-coded bits, each user decodes the information bits of all the other users by applying an iterative message-passing decoder \cite{LT}. The transmission continues until all users have successfully received all information bits from all other users.
The resulting performance is measured by the overhead defined as $\varepsilon=N_r/(r-1)K$, where $N_r$ is the number of correctly received coded bits on successful decoding. Another useful measure is the transmission overhead defined as $\varepsilon_t=N_{\text{max}}/(r-1)K$, where $N_{\text{max}}$ is the maximum number of transmitted code bits by any user at the instance of successful decoding by all users.
\section{Performance Analysis}
\subsection{Lossless Uplink Phase}
For simplicity, we first consider a network with lossless uplink channels.
It follows that $\epsilon_{u_i}=0$ and $\alpha_i=1$ for $i=1,2,\ldots,r$. We assume that $\Omega(x)=x$; hence each user transmits a randomly chosen information bit to the relay in the uplink. Also, we assume symmetric links in the broadcast phase such that $\epsilon_{d_i}=\epsilon_{d}$ for all $i$. Consequently, the value of a relay-check node in the $j$-th broadcast phase is computed as
\begin{equation}\label{}
x_j=\overset{D}{\underset{i=1}{\bigoplus}}\psi_i c_i,
\end{equation}
and broadcast to all users. Before decoding starts at user $i$, the information bits of user $i$, encoded by the relay, are removed through modulo-2 addition.
 The check-node degree distribution observed for decoding of user $i$ is therefore modified as 
\begin{eqnarray}\label{dnDD}
\Phi_d &=&\sum_{k=0}^{1}\Gamma_{d+k}\frac{\dbinom{K}{k} \dbinom{r-1}{d}\dbinom{K}{1}^{d}}{\sum_{i=0}^{1}\dbinom{K}{i}\dbinom{r-1}{d+k-i}\dbinom{K}{1}^{d+k-i}} \nonumber \\
       &=& \sum_{j=d}^{d+1}\Gamma_j \frac{\dbinom{r-1}{d}}{\dbinom{r}{j}},
\end{eqnarray}
where $d=0,1,\ldots,D-1$. Hence, a coded bit of degree $d$ for the decoding of user $i$ will have been of degree $d+k$ with probability $\Gamma_{d+k}$ before the actual information bits transmitted by user $i$ were removed. Furthermore, we observe that $\Phi_0\neq 0$, indicating that after the removal of all known information bits 
there may exist some check nodes, which are not connected to any variable nodes in the decoding graph. As these check nodes are useless for decoding they are removed, 
and the resulting check-node degree distribution is normalized as
\begin{equation}\label{norm}
\hat{\Phi}_d=\frac{\Phi_d}{(1-\Phi_0)}, d=1,2,\ldots,D-1.
\end{equation}
The expression in \eqref{dnDD} can be transformed into the 
edge-perspective form as follows
\begin{equation}\label{deDD}
\phi_d=\sum_{j=d}^{d+1}\gamma_{j}\frac{d \Gamma'(1) \dbinom{r-1}{d}}{j\Phi'(1)\dbinom{r}{j}}, d=0,1,\ldots,D-1.
\end{equation}
The probability that a variable node of user $i$ is not recovered after $\ell$ decoding iterations is asymptotically given by
\begin{equation}\label{DE}
P_{i,\ell} \approx e^{-\bar{\mu}\phi(1-P_{i,\ell-1})},
\end{equation}
where $\bar{\mu}$ is the average variable-node degree after removing the known variable nodes from the decoding graph. The overhead can be expressed in terms of $\sum_j \phi_j/j$. 
We can therefore formulate a linear program for the optimization of the check edge-perspective degree distribution at an arbitrary user given lossless uplink channel as follows:
\begin{align*}\label{eqn:linprog1}
\text{LP1}:\quad \quad \text{min}\quad  \sum_{j=1}^{D-1} \phi_j/j \quad \quad \quad&\\
\sum_{j=1}^{D-1}\phi_{j}x_n^{j-1}\geq-\frac{-\ln(1-x_n)}{\bar{\mu}}&,
\end{align*}
where $n\in 1,2,\ldots,m$ and $0=x_1<x_2<\ldots<x_m=1-\delta$ are equidistant points on $[0, 1-\delta]$ and $\delta$ is the desired erasure rate. LP1 is solved for a series of values of $\bar{\mu}$ to determine the optimal check edge-perspective degree distribution $\phi(x)$  in terms of overhead. Unfortunately it is difficult to track the relay check edge-perspective degree distribution from LP1. We can further transform LP1 to obtain the relay check edge-perspective degree distribution leading to the required optimized degree distribution at the user decoders. For this aim, we express the overhead in terms of relay edge-perspective degrees as $\varepsilon=\bar{\gamma}\sum_{d=1}^{D}\gamma_d/d$ where $\bar{\gamma}$ is the average variable-node degree before the removal of the known variable nodes from the decoding graph. Similarly, the constraints in LP1 can also be transformed in terms of $\gamma_d$ by substitution of Eqn. \eqref{deDD}. The modified linear program is given by
\begin{align*}
\text{LP2}:&\quad \text{min}\quad \sum_{d=1}^{D} \gamma_d/d \quad \quad \quad \quad \quad \quad& \\
&\sum_{d=1}^{D}\sum_{j=d}^{d+1}\gamma_{j}\frac{d\dbinom{r-1}{d}}{j\dbinom{r}{j}}x_n^{d-1}\geq-\frac{-\ln(1-x_n)}{\bar{\gamma}}.&\\
\end{align*}
As an example, we obtain the relay check edge-perspective degree distribution by running LP2 for a 10-user scenario with symmetric downlink channels and $\delta=0.02$. The corresponding node-perspective degree distribution is given by
\begin{equation}\label{RDD}
\Gamma(x)=0.0058x+0.4281x^2+0.3411x^3+0.2250x^{10}.
\end{equation}
The degree distribution at the decoder for any user can be obtained by using Eqn. \eqref{dnDD} and Eqn. \eqref{norm}. Since the inherent structure of an LT code is identical to a non-systematic low-density generator matrix (LDGM) code, our construction has generically an erasure floor. We note that our proposed DLT codes suffer from high erasure floor for $r\leq3$. However, the erasure floor can be mitigated by concatenating a high-rate pre-code at the users in our proposed DLT codes similar to Raptor codes \cite{Sh06}.


\subsection{Lossy Uplink Phase}
The design framework detailed in LP2 optimizes the degree distributions in terms of minimizing reception overhead. However, the delay due to lossy uplink transmission is not considered in this optimization process. Consequently, a higher transmission overhead is required as compared to lossless uplink transmission. 
When the uplink channels are lossy, some of the coded bits from the users will be erased. The relay, however, still randomly chooses $d (\leq D)$ users to generate a relay-coded bit as detailed in Eqn. \eqref{RCB}. In case one of the selected bits is erased, the relay is not able to complete the proper encoding process.  Some alternatives are to apply Eqn. \eqref{RCB} regardless of possible erasures; to select only from received bits; or to halt downlink transmission and repeat uplink transmissions until all bits are received similar to \cite{SPD09}. The first two alternatives result in inferior coding, while the latter alternative introduces delay.

To address this problem, we propose to use a one-bit buffer at the relay for each user-relay uplink connection. In this case the relay stores the coded bit from user $i$ in buffer $B_i$. Each buffer is updated on the reception of a new coded bit from the corresponding user by overwriting the previously stored coded bit with a newly received coded bit. Yet, at the beginning of transmission, the relay must wait until all buffers are loaded; after which the relay can transmit properly encoded bits in each broadcast phase without any further delay. If a coded bit from user $i$ is erased in the current uplink phase then the relay will use the previously stored coded bit in buffer $B_i$ in the broadcast phase. Some additional control information is required to keep track of exactly what user bits are encoded in a particular transmission. The initial delay required for loading the buffers is negligible since $\epsilon_{u_i}$ is typically small for $i=1,2,\ldots,r$ and the number of information bits at each source is large. The relay-coded bit in the $j$-th broadcast phase can be determined as
\vspace{-2mm}
\begin{equation}\label{}
x_j=\overset{D}{\underset{i=1}{\bigoplus}}\psi_i b_i,
\end{equation}
where $b_i$ is the coded bit of user $i$ stored in buffer $B_i$ in the $j$-th broadcast phase. 
\vspace{-2mm}
\section{Numerical Examples}
\begin{figure}[!t]
\centering
\includegraphics[width=68mm]{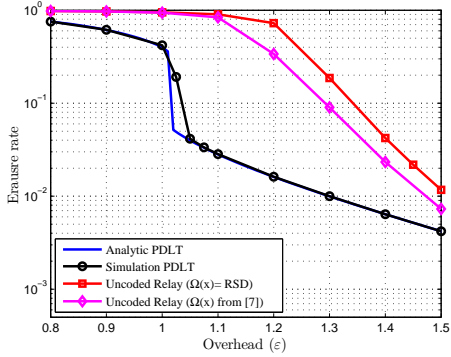}
\caption{Performance for lossless uplink transmission.}
\vspace{-5mm}
\label{Sim_1ossless}
\end{figure}
Here we present numerical results to highlight the erasure rate performance improvements provided by of our proposed DLT (PDLT) codes. We consider a network with 10 users each having $K=1000$ information bits, and assume ideal uplink channels such that $\epsilon_{u_i}=0$ for $i=1,2,\ldots,r$. We further assume symmetric downlink channels with erasure probabilities $\epsilon_{d_1}=\epsilon_{d_2}=\ldots=\epsilon_{d_r}=0.1$. The check node degree distribution at the users is $\Omega(x)=x$, while the relay check-node degree distribution is detailed in Eqn. \eqref{RDD}. The simulated performance and the corresponding asymptotic performance based on Eqn. \eqref{DE} are depicted in Fig.~\ref{Sim_1ossless}. The results clearly demonstrates a close match of the finite-length example to the asymptotic performance. We also provide a comparison to the case where the relay simply broadcasts instead of combining the coded bits received from each user in its allocated time slots in uplink transmission. In the uncoded scenario, decoding at each user is performed on $(r-1)K$ separate decoding graphs each containing corresponding other user variable and check nodes. For this scenario we consider two different degree distributions at the users. In the first case, we use the robust Soliton distribution (RSD) with corresponding parameters $K=1000$, $c=0.03$ and $\sigma=0.05$ \cite{LT} to encode the information bits at users. In the second case we use the degree distribution specified in \cite{Sh06}. From Fig.~\ref{Sim_1ossless}, we can observe that our proposed DLT code has better performance than the two uncoded relay scenarios.

\begin{figure}[!t]
\centering
\includegraphics[width=68mm]{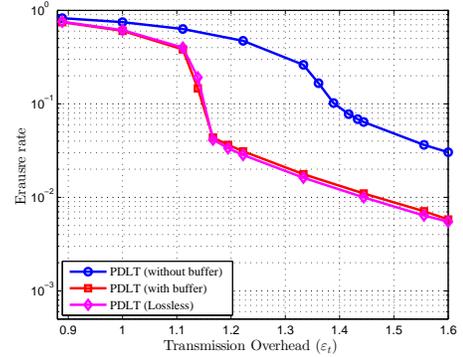}
\caption{Performance for lossy uplink transmission.}
\label{Sim_lossy}
\vspace{-5mm}
\end{figure}

For a more realistic EMR channel, we consider symmetric lossy uplink channels with erasure probabilities $\epsilon_{u_1}=\epsilon_{u_2}=\ldots=\epsilon_{u_r}=0.05$. The remaining parameters are the same as for the lossless case. A comparison of our proposed DLT codes with  and without buffer is presented Fig.~\ref{Sim_lossy}. Here we can easily observe that our proposed DLT code with buffer comprehensively outperforms its counterpart. It is also demonstrated in Fig.~\ref{Sim_lossy} that our DLT coding scheme has virtually no loss of performance when comparing the cases of lossless and lossy uplink transmission. The approach can readily be extended to asymmetric channels and different values of channel erasure probabilities. 

\vspace{-2mm}
\section{Conclusions}
We have developed a design framework for DLT codes intended for transmission over the EMR channel. The framework is based on density evolution, and allows us to obtain an optimized degree distribution at the relay in terms of minimizing the overhead. We have further proposed a buffer-based strategy, such that the relay can transmit properly encoded packets regardless of potential packet losses in the uplink transmissions. The performance of the proposed codes are evaluated through simulations and found to match perfectly with the asymptotic performance determined by density evolution. Finally our results demonstrated that the proposed buffer-based DLT codes comprehensively outperform its counterparts for lossy uplink channels in multi-way relay networks.

\bibliographystyle{IEEEtran}
\bibliography{IEEEabrv,references}

\end{document}